\documentclass[pre,aps,twocolumn,superscriptaddress,floatfix]{revtex4}

\usepackage{graphicx}
\usepackage{amsmath,empheq}
\usepackage{amssymb}
\usepackage{ifthen}
\usepackage{booktabs}

\usepackage{graphicx}
\usepackage{dcolumn}
\usepackage{bm}
\usepackage{gensymb}
\usepackage{color}

\newcommand{\bb}{\begin{equation}}
\newcommand{\ee}{\end{equation}}
\newcommand{\ba}{\begin{eqnarray*}}
\newcommand{\ea}{\end{eqnarray*}}
\newcommand{\rhor}{\rho({\bf r})}
\newcommand{\dd}{{\rm d}}
\newcommand{\rr}{{\mathbf r}}
\newcommand{\xx}{{\mathbf x}}
\newcommand{\dr}{{\rm d}{\bf r}}

\newcommand{\RR}{{\mathcal{R}}}

\makeatletter
\def\@frameeq#1{%
  \framebox{$\,\displaystyle#1\hbox{\vrule height 2.4ex depth 1.4ex width 0pt}\,$}}
\newcommand\Equation[1]{$$\refstepcounter{equation}%
  \@frameeq{#1}%
  \eqno \hbox{\@eqnnum}$$\@ignoretrue\ignorespaces}
\newcommand\Displaystyle[1]{$$\@frameeq{#1}$$\@ignoretrue\ignorespaces}
\makeatother

\usepackage{longtable}

\newcounter{subfigcount}
\newcounter{figcount}
\newcommand{\subfloat}[3]{%
{\ifnum\thefigure=\thefigcount\stepcounter{subfigcount}%
\else\setcounter{figcount}{\thefigure}\setcounter{subfigcount}{1}\fi%
}%
\noindent%
\begin{minipage}[b]{#1}%
  \centering%
  {#3}\\[0pt]%
  (\alph{subfigcount})~#2%
\end{minipage}}%

\newcommand{\subfloatflex}[2]{%
{\ifnum\thefigure=\thefigcount\stepcounter{subfigcount}
\else\setcounter{figcount}{\thefigure}\setcounter{subfigcount}{1}\fi%
}%
\noindent%
\begin{minipage}[b]{\widthof{#2}}%
  \centering%
  {#2}\\[0pt]%
  (\alph{subfigcount})~#1%
\end{minipage}}%

\begin{document}

 \normalsize

\title{Complete Wetting and Drying at Sinusoidal Walls}

\author{Alexandr \surname{Malijevsk\'y}\footnote{malijevsky@icpf.cas.cz}}
\affiliation{Research group of Molecular and Mesoscopic Modelling, The Czech Academy of Sciences, Institute of Chemical Process Fundamentals, 165 02
Prague, Czech
  Republic}
\affiliation{Department of Physical Chemistry, University of Chemistry and Technology Prague, 166 28 Prague, Czech Republic}

\author{Martin \surname{Posp\'\i\v sil}}
\affiliation{Research group of Molecular and Mesoscopic Modelling, The Czech Academy of Sciences, Institute of Chemical Process Fundamentals, 165 02
Prague, Czech
  Republic}

\author{Miriam  \surname{Mago\v ciov\'a}}
\affiliation{Department of Physical Chemistry, University of Chemistry and Technology Prague, 166 28 Prague, Czech Republic}

\author{Ji\v r\'\i \hspace{0.001cm} \surname{Janek}}
\affiliation{Research group of Molecular and Mesoscopic Modelling, The Czech Academy of Sciences, Institute of Chemical Process Fundamentals, 165 02
Prague, Czech Republic}

\begin{abstract}
\noindent We investigate complete wetting and drying at sinusoidally corrugated solid walls, focusing on the effects of wall geometry and interaction
range. Two distinct interaction models are considered: one incorporating only short-ranged (SR) forces (applied to drying), and another including
long-ranged (LR) van der Waals interactions (applied to wetting). The SR model is analyzed within the framework of nonlocal Hamiltonian theory by
Parry et al., while the LR model is treated using a sharp-kink approximation. In both cases, we derive scaling relations that describe the dependence
of the adsorbed layer's width and morphology on the wall's geometric parameters as the system approaches two-phase coexistence. We identify distinct
scaling regimes determined by the degree of wall corrugation and highlight the contrasting effects of SR and LR interactions. Theoretical predictions
are corroborated by numerical results from classical density functional theory.
\end{abstract}

\maketitle

\section{Introduction}

Consider a simple one-component fluid exhibiting liquid-gas phase separation below its critical temperature, $T_c$, which is brought into contact
with a solid planar wall. According to Cahn's argument \cite{cahn}, there exists a wetting temperature, $T_w<T_c$, at which Young's contact angle
$\theta$ of a macroscopic liquid droplet at the wall vanishes and which thus separates partially wet states ($\theta>0$ for $T<T_w$) from completely
wet states ($\theta=0$ for $T\ge T_w$). Although Cahn's argument has been shown to be generally incorrect -- its validity depends on the range of
microscopic forces \cite{degennes, joseph85, esw, joseph22, parry23, parry24, parry24b} -- it introduced a novel class of  phase transitions
associated with singularities in the surface contribution to the system free energy. The simplest example of this is the wetting transition occurring
on a two-phase coexistence line as $T_w$ ia approached from below, whose order depends sensitively on the nature of the microscopic interactions
\cite{croxton, sull, dietrich, schick, forgacs}.

Alternatively, the wetting transition may be characterized by a divergence of the thickness of the adsorbed liquid layer intruding the wall-gas
interface. This also allows for a description of associated phenomena such as \emph{complete wetting}, the continuous phase transition obtained as
the bulk coexistence is approached from below along an isotherm $T>T_w$. For planar walls, this critical phenomenon is characterized by the power-law
divergences of the wetting film thickness $\ell_\pi$ and the parallel correlation length $\xi_{\parallel}^{\rm co}$:
 \bb
 \ell_\pi\sim\delta\mu^{-\beta^{\rm co}}\,,\;\;\;\xi_{\parallel}^{\rm co}\sim\delta\mu^{-\nu_\parallel^{\rm co}}\,, \label{ell_pi}
 \ee
as $\delta\mu\equiv|\mu_{\rm sat}(T)-\mu|$, the deviation of the chemical potential from the saturation, tends to zero. In three-dimensional systems,
the critical exponents $\beta^{\rm co}$ and $\nu_\parallel^{\rm co}$ depend on the range of intermolecular interactions: in particular, for systems
with dispersion forces, $\beta^{\rm co}=1/3$ and $\nu_\parallel^{\rm co}=2/3$, while for those governed solely by short-range interactions,
$\beta^{\rm co}=0(\ln)$ and $\nu_\parallel^{\rm co}=1/2$.

At weakly attractive (or fully repulsive) walls, a closely related phenomenon of \emph{complete drying} may occur. This refers to the process when
two-phase coexistence is approached from above (i.e., from the liquid side), leading to the adsorption of a gas layer. In the limit $\delta\mu\to0$,
the thickness of this adsorbed layer diverges according to the same power-law behaviour as in Eq.~(\ref{ell_pi}). For fluids, the only asymmetry
between complete wetting and drying arises from molecular packing effects near the wall, which are significantly more pronounced when the adsorbed
phase is liquid.

The range of possible interfacial phenomena extends significantly when the walls are no longer smooth. Experimental studies have shown
\cite{bruschi02,gang,bruschi06}  that structured walls may induce novel types of interfacial phase transitions absent in the planar case, with
critical exponents strongly dependent on the wall geometry \cite{nature}. For instance, when a wall is patterned with grooves or pits, complete
wetting is preceded by a filling transition -- which is either abrupt or continuous -- followed by a depinning transition, where the liquid-gas interface
detaches from the wall surface \cite{darbellay,robbins,tas06,tas07,hofmann,mal13,rascon13,our_groove,singh,mal20,singh22,oktasendra}.

 In this work, we concentrate on another very simple model characterizing wall corrugation, assuming that the wall shape varies in a sinusoidal
manner. This model has been the subject of intense research investigating the impact of wall undulations on the phase boundary of the wetting
transition, its order, and the presence of an unbending transition characterized by an abrupt flattening of the liquid-gas interface
\cite{fox,netz,unbend,rascon2000,rejmer2000,rejmer02,rejmer07,patricio}. Additional mesophase transitions may also occur in complex fluids
\cite{patricio1, patricio2, patricio3, rojas}. More recently, a continuous phase transition referred to as osculation has been predicted for
sinusoidal walls above the wetting transition. This transition occurs at the pressure where the corresponding Laplace curvature matches the geometric
curvature of the substrate troughs \cite{osc,osc2}, marking the onset of macroscopic adsorption governed by wall geometry.

Here, we focus on the ultimate adsorption regime of sinusoidally corrugated walls, which occurs in the immediate vicinity of fluid saturation when
$\delta\mu \to 0$. The walls are assumed to be either completely wet or completely dry, and our aim is to provide a detailed description of how wall
corrugation influences the process of complete wetting or drying. To this end, a microscopic approach that accounts for molecular forces is required.
This enables us to contrast the qualitatively distinct wetting behavior of systems with long-range (LR) versus short-range (SR) interactions. Recall
that, in the context of wetting phenomena, LR interactions decay as an inverse power law, while SR interactions are either of finite range or their
 asymptotic decay is at least exponentially fast.

To this end, we adopt two sinusoidally shaped substrate models:
i) a substrate formed of uniformly distributed Lennard-Jones particles representing a LR potential, which exhibits complete wetting above its wetting
temperature; and
ii) a purely repulsive substrate modelled by a simple hard wall, representing a SR potential, which exhibits complete drying at all temperatures.

For both substrate models, our main objective is to describe the scaling behavior of the complete wetting/drying transition in terms of the
properties of the liquid-gas interface, $\ell(x)$ -- specifically, its mean height and shape -- as functions of the wall parameters: the amplitude
$A$ and the period $L$ (or equivalently, the wave number $k = 2\pi/L$), in the limit $\delta\mu \to 0$.

The paper is organized as follows: In the next section, we introduce the microscopic models used to describe fluid adsorption on sinusoidal walls,
defining the fluid-fluid and wall-fluid interactions. We also formulate the nonlocal density functional theory (DFT) framework within which these
systems are studied at the microscopic level. The following Section III focuses on a coarse-grained approach, describing the system in terms of the
interface height and shape, that allows for analytical approximations. This section is divided into two parts: one presenting a nonlocal interfacial
Hamiltonian theory appropriate for a description of interfacial phenomena of SR systems at non-planar geometries, and the other devoted to an
analysis of LR systems by employing the sharp-kink approximation. In both cases, we derive the asymptotic properties of the mean height and
corrugation of the unbinding liquid-gas interface in terms of their scaling with $\delta\mu$ and the geometric parameters of the wall. In Section IV,
we test these predictions by comparing with the numerical results obtained from DFT. Finally, Section V summarizes the main findings of this work and
discusses its possible generalizations.

\section{Microscopic model}

In this section, we formulate a microscopic model of a single, one-component fluid which is at contact with  a sinusoidally shaped wall exerting an
external field $V(\rr)$ and at equilibrium  with a bulk reservoir at temperature $T$ and chemical potential $\mu$. Within classical density
functional theory (DFT) \cite{evans79}, the equilibrium density profile $\rho(\rr)$ is determined by minimizing the grand potential functional
   \bb
  \Omega[\rho] =F[\rho]+\int\dr\rho(\rr)\left[V(\rr)-\mu\right]\,,\label{grandpot}
 \ee
where $F[\rho]$ represents the intrinsic free energy functional, which contains all inter-particle interactions. The choice of approximation for
$F[\rho]$ is crucial in any DFT model and strongly depends on the specific fluid system. For simple fluids, a common perturbative approach decomposes
$F[\rho]$ into three contributions
 \bb
 F[\rho]=F_{\rm id}[\rho]+F_{\rm hs}[\rho]+F_{\rm att}[\rho]\,, \label{f_dft}
 \ee
where $F_{\rm id}$ is the ideal gas term, $F_{\rm hs}$ accounts for repulsive interactions and $F_{\rm att}$ describes attractive interactions.

The ideal gas contribution is known exactly and is given by
  \bb
  \beta F_{\rm id}[\rho]=\int\dr\rho(\rr)\left[\ln(\rhor\Lambda^3)-1\right]\,,
  \ee
where $\Lambda$ is the thermal de Broglie wavelength (which can be set to unity) and $\beta=1/k_BT$ is the inverse temperature.

The repulsive interactions between fluid molecules are modelled using the hard-sphere potential, with the corresponding free energy contribution
accurately described by Rosenfeld's fundamental measure theory (FMT) \cite{ros}
 \bb
F_{\rm hs}[\rho]=\frac{1}{\beta}\int\dd\rr\,\Phi(\{n_\alpha\})\,,\label{fmt}
 \ee
 where $\{n_\alpha\}$ is a set of six weighted densities defined as
 \bb
 n_\alpha(\rr)=\int\dr'\rho(\rr')\omega_\alpha(\rr-\rr')\,,\;\;\alpha=\{0,1,2,3,v_1,v_2\}\,. \label{n_alpha}
 \ee
The weighted densities are obtained by convoluting the one-body density $\rhor$ with the weight functions $\{\omega_\alpha\}$, which correspond to
the fundamental geometric measures of hard spheres with diameter $\sigma$:
 \begin{eqnarray}
 \omega_3(\rr)&=&\Theta(\RR-r)\,,\;\;\;\;\;\omega_2(\rr)=\delta(\RR-r)\,,\\
 \omega_1(\rr)&=&\omega_2(\rr)/4\pi \RR\,,\;\;\;\omega_0(\rr)=\omega_2(\rr)/4\pi \RR^2\,,\\
 \omega_{v_2}(\rr)&=&\frac{\rr}{\RR}\delta(\RR-r)\,,\;\omega_{v_1}(\rr)=\omega_{v_2}(\rr)/4\pi \RR\,.
 \end{eqnarray}
Here, $\Theta$ denotes the Heaviside function, $\delta$ is Dirac's delta function, $r=|\rr|$ and $\RR=\sigma/2$. Among the various prescriptions for
the free energy density $\Phi$ of an inhomogeneous hard-sphere fluid, we employ the original Rosenfeld approximation \cite{ros}, which accurately
captures short-range correlations between fluid particles while satisfying exact statistical mechanical sum rules \cite{hend}.

\begin{figure}[htbp]
  \includegraphics[width=\linewidth]{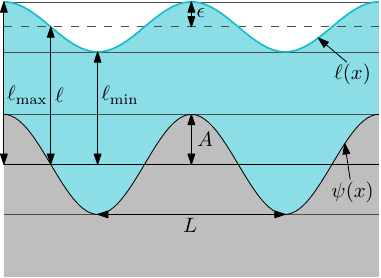}
  \caption{A schematic of our model showing a wall with a local height profile $\psi(x)$ (relative to the horizontal, indicated by the thick solid line), which varies smoothly with period $L$ and has amplitude $A$. An adsorbed wetting layer of local height $\ell(x)$ is also depicted. The mean height of the layer above the wall is $\ell$, as defined in Eq.~(\ref{ell_dft}), and its corrugation is characterized by the parameter $\epsilon$, defined in Eq.~(\ref{eps_dft}). }
  \label{scheme}
\end{figure}

For particle separations $r>\sigma$,  the interaction between fluid particles is governed by the attractive part of the Lennard-Jones potential,
$u_{\rm att}(r)$, which is truncated at a cut-off distance $r_c=2.5\,\sigma$:
 \bb
 u_{\rm att}(r)=\left\{\begin{array}{cc}
 0\,;&r<\sigma\,,\\
-4\varepsilon\left(\frac{\sigma}{r}\right)^6\,;& \sigma<r<r_c\,,\\
0\,;&r>r_c\,.
\end{array}\right.\label{uatt}
 \ee
This attractive contribution is included in the free-energy functional using a simple mean-field approximation
 \bb
F_{\rm att}[\rho]=\frac{1}{2}\int\int\dd\rr\dd\rr'\rhor\rho(\rr')u_{\rm att}(|\rr-\rr'|)\,.
 \ee

Two different wall potentials are considered. In the first model, the wall acts as an impenetrable barrier described by a hard-wall potential:
 \bb
 V_d(x,z) = \left\{
  \begin{array}{ll}
  \infty, & z< \psi(x)\,, \\
   0, & z > \psi(x)\,,
  \end{array} \right.
 \ee
where the function
 \bb
 \psi(x)=A\cos(kx) \label{psi}
  \ee
defines the local height of the wall above the horizontal plane at $z=0$. In this purely repulsive substrate model the contact angle is $\theta=\pi$
at any subcritical temperature and is thus suitable for a DFT analysis of complete drying.

The second wall model  $V_w(z)$  arises from a uniform distribution of wall atoms with number density $\rho_w$, which interact with fluid atoms via
the Lennard-Jones potential
 \bb
 \phi_w(r)=4\varepsilon_w\left[\left(\frac{\sigma}{r}\right)^{12}-\left(\frac{\sigma}{r}\right)^6\right]\,. \label{r6}
 \ee
The net wall potential is obtained by integrating $\phi_w(r)$ over the entire volume of the wall
  \bb
 V_w(z)=
\begin{cases}
   \rho_w\int_{z<\psi(x)}\dr\phi_w(r)\,, & z>\psi(x)\,,\\
  \infty\,, & z<\psi(x)\,, \label{Vw}
\end{cases}
 \ee
decaying asymptotically as $z^{-3}$.

The inclusion of dispersion interactions through $V_w(z)$ implies that the second model describes long-ranged systems, characterized by algebraically
decaying microscopic interactions.  This model is used to study complete wetting at a wall-gas interface, following the path $\mu\to\mu_{\rm sat}^-$
along an isotherm with $T>T_w$, where $T_w$ is the wetting temperature of the corresponding planar wall. Throughout this study we consider the wall
interaction parameter $\varepsilon_w=\varepsilon$, for which  $T_w=0.8\,T_c$  with $T_c\approx1.41\,\varepsilon/k_B$ being the bulk critical
temperature. In contrast, the first model includes only short-ranged  interactions, which decay exponentially or have a finite range. It is used to
study complete drying at a wall-liquid interface, following the path   $\mu\to\mu_{\rm sat}^+$ along the same temperature.

Minimizing Eq.~(\ref{grandpot}) yields the self-consistent equation for the equilibrium density profile
  \bb
  \rho(\rr) = \Lambda^{-3} \exp\left[\beta\mu-\beta V(\rr) + c^{(1)}(\rr)\right]\,,  \label{selfconsistent}
 \ee
where $c^{(1)}(\rr)=c^{(1)}_\mathrm{hs}(\rr)+c^{(1)}_\mathrm{att}(\rr)$ is the one-body direct correlation function, which can be split into the
hard-sphere contribution
 \bb
  c^{(1)}_\mathrm{hs}(\rr)  
   = -\sum_\alpha \int\dd\rr'\; \frac{\partial\Phi(\{n_\alpha\})}{\partial n_\alpha} \, \omega_\alpha(\rr'-\rr)   \label{chs}
 \ee
and the attractive part
  \bb
  c^{(1)}_\mathrm{att}(\rr) 
  =-\beta\int \dd\rr'\; u_{\rm att}(|\rr-\rr'|)\,\rho(\rr')\,.\label{catt}
  \ee

We solve Eq.~(\ref{selfconsistent}) numerically on a discrete rectangular grid with a spacing of $0.1\,\sigma$ using the Picard iteration. The
convolutions in Eqs.~(\ref{n_alpha}), (\ref{chs}), and (\ref{catt}) are computed via fast Fourier transform (see Ref.~\cite{posp22} for details). To
this end, we impose the periodic boundary conditions along the $x$-axis and assume that the density at the upper boundary of our computational domain
is equal to the bulk fluid density.

From the equilibrium density profile $\rho(x,z)$, we  determine the local height of the liquid-gas interface using a standard crossing criterion:
 \bb
 \rho(x,\ell(x))=\frac{\rho_g+\rho_l}{2}\,, \label{dft_cont}
 \ee
where the coexisting  bulk densities $\rho_g$ and $\rho_l$ of the gas and liquid phases, respectively, are  obtained from Eq.~(\ref{selfconsistent})
by setting $V(\rr)=0$.

Once the interface profile $\ell(x)$ is determined, we define the mean interface height as
 \bb
 \ell=\frac{\ell_{\rm max}+\ell_{\rm min}}{2}  \label{ell_dft}
 \ee
and the corrugation magnitude as
  \bb
  \epsilon=\frac{\ell_{\rm max}-\ell_{\rm min}}{2}  \label{eps_dft}
  \ee
  where $\ell_{\rm max}$ and $\ell_{\rm min}$ are the maximum and minimum interface heights, respectively. These characteristics are illustrated in Fig.~1, which shows a schematic of our model.

\section{Mesoscopic approach}

In this section, we introduce coarse-grained models allowing for approximative analytic predictions. Unlike the molecular-scale treatment presented
above, these models integrate out microscopic degrees of freedom and describe the system's behavior in terms of the local height of the liquid-gas
interface, $\ell(\xx)$. The section is split into three parts. We begin by briefly recalling the interfacial Hamiltonian theory of complete wetting
for a planar wall.  The second part focuses on the nonlocal Hamiltonian theory, which is suitable for SR interactions and will be applied for an
analysis of complete drying. Finally, the third part is dedicated to the sharp-kink approximation, which will be used for a description of complete
wetting in LR systems.

\subsection{Complete wetting on a planar wall}

The interfacial Hamiltonian approach has been widely used to study interfacial phenomena on  planar walls, particularly for determining critical
exponents and their dependence on the range of intermolecular forces. In three dimensions, the interfacial Hamiltonian is typically of the form
 \bb
H_\pi[\ell]=\int d{\bf x}\left\{\frac{\gamma}{2}(\nabla\ell(\xx))^2+W_\pi(\ell)\right\}\,,  \label{Hpi}
 \ee
where ${\bf x}=(x,y)$ are  coordinates parallel to the wall, which occupies the region $z<0$. The first term accounts for the distortion of the
interface due to thermal fluctuations, treated  within the square-gradient approximation. The second term, $W_\pi(\ell)$,  represents the binding
potential, which involves the effective interaction between the interface with the planar wall arising from intermolecular forces.

For SR interactions, the binding potential takes the form
 \bb
  W_\pi(\ell)=\delta\mu\Delta\rho\ell+a(T){\rm e}^{-\kappa\ell}+\cdots\,, \label{vsr}
 \ee
together with a hard-wall repulsion restricting the interface height to $\ell(\xx)>0$. Here, $\kappa=\xi_b^{-1}$ is the inverse of the bulk
correlation length, which characterize the liquid-gas interface width and $\Delta\rho=\rho_l-\rho_g$.

In systems with LR interactions, the binding potential  exhibits an algebraic decay
 \bb
  W_\pi(\ell)=\delta\mu\Delta\rho\ell+\frac{A_H(T)}{\ell^p}+\cdots\,, \label{vlr}
 \ee
where a hard-wall repulsion at $z=0$ is again assumed. This includes systems with dispersion (van der Walls) forces, for which $p=2$ and $A_H$ is the
Hamaker constant.

In three dimensions, the role of interfacial fluctuation is unimportant for complete wetting (drying), hence the equilibrium mean height of the
interface, $\ell_\pi$, can be obtained simply by minimizing the binding potential:
 \bb
  \left.\frac{dW_\pi(\ell)}{d\ell}\right|_{\ell=\ell_\pi}=0\,,  \label{mf}
 \ee
This leads to the following asymptotic expressions for SR systems
  \bb
  \ell_\pi\approx-\xi_b\ln(\delta\mu)\;\;\;({\rm SR})\,,
  \ee
corresponding to $\beta_{\rm co}=0(\ln)$), and
   \bb
    \ell_\pi\approx \left(\frac{2A_H}{\delta\mu}\right)^{\frac{1}{3}} \;\;\;({\rm LR})\,,
   \ee
 for LR systems with dispersion forces, which corresponds to $\beta_{\rm co}=1/3$.

\subsection{Nonlocal Hamiltonian theory} \label{ssec_nonlocal}

Within the nonlocal Hamiltonian model, the energy cost of a given interfacial configuration is described by the functional \cite{nonlocal}
 \begin{eqnarray}
H[\ell]&=&\int d{\bf x}\left\{\gamma\sqrt{1+(\nabla\ell(\xx))^2}+\delta\mu[\ell(\xx)-\psi(\xx)]\right\}\nonumber\\
&&+W[\ell,\psi]\,. \label{ham}
 \end{eqnarray}
Here, the free-energy cost due to the liquid-gas interface is treated within the drumhead model and the contribution  accounting for the presence of
the metastable liquid is singled out. The final term is the nonlocal binding potential functional, which is appropriate for describing wall and/or
interface corrugations. It is given by the expansion
 \bb
 W[\ell,\psi]=\sum_{n=1}^\infty\left(a_1\Omega^n_n[\ell,\psi]+b_1\Omega^{n+1}_n[\ell,\psi]+b_2\Omega^n_{n+1}[\ell,\psi]\right)\,, \label{nl}
 \ee
with the coefficients $a_1=\sqrt{8\gamma_{wl}\gamma}$, $b_1=\gamma$, and $b_2=2\gamma_{wl}$. The functionals  $\Omega_i^j[\ell,\psi]$ represent
multiple integrals that couple $i$ points on the wall to $j$ points on the interface, each pair connected by the kernel $K(\rr_1,\rr_2)$, a two-point
function defined as:
 \bb
 K(\rr_1,\rr_2)=\frac{\kappa}{2\pi}\frac{e^{-\kappa |\rr_2-\rr_1|}}{|\rr_2-\rr_1|}\,, \label{K}
 \ee
which is the Green function of the  Helmholtz differential equation associated with the underlying Landau--Ginzburg--Wilson model, solved within the
double-parabola approximation \cite{nonlocal} and corresponds to the bulk three-dimensional correlation function \cite{schick}.

For complete wetting/drying it is sufficient to consider only the first term in the expansion (\ref{nl}), which is of the form
 \bb
\Omega^1_1[\ell,\psi]=\int d{\bf s}_\psi \int d{\bf s}_\ell K(\rr_\psi,\rr_\ell) \label{omega11}
 \ee
where $\xx=(x,y)$ is the parallel displacement vector, $\rr_\psi=(\xx,\psi(\xx))$ and $\rr_\ell=(\xx,\ell(\xx))$ are the points at the wall and the
interface, respectively, and $d{\bf s}_\psi=\sqrt{1+(\nabla\psi)^2}d\xx$, $d{\bf s}_\ell=\sqrt{1+(\nabla\ell)^2}d\xx$ are the corresponding surface
elements.

On a mean-field level, the equilibrium interfacial profile is obtained by solving the Euler-Lagrange (EL) equation
 \bb
 \left.\frac{\delta H}{\delta \ell}\right|_{\rm eq}=0\,, \label{EL}
 \ee
 which can be accomplished fully numerically or using various approximations discussed below.

\subsubsection{Flat-interface approximation}

We begin by considering the simplest approximation, in which the liquid-gas interface is assumed to be flat, $\ell(x)=\ell$, and focus on the scaling
behavior of $\ell$ with respect to the geometric parameters of the wall. The governing equations are first formulated in general for a wall with
one-dimensional periodic corrugation, $\psi(x+L)=\psi(x)$, before specializing to the sinusoidal case.

Exploiting the symmetry of the system, the interfacial Hamiltonian per unit length and per period reduces to
 \bb
H=\delta \mu\Delta\rho\int_0^L dx \left[\ell-\psi(x)\right]+a_1\Omega_1^1+\cdots\,,  \label{Hflat}
 \ee
where
\begin{eqnarray}
\Omega^1_1&=&\int_0^L dx\sqrt{1+\psi'(x)^2}e^{-\kappa(\ell-\psi(x))}\nonumber\\
&=&Le^{-\kappa(\ell-\psi_{\rm max})}g\,.
\end{eqnarray}
 Here, the dimensionless auxiliary function $g$ defined as
 \bb
 g=\frac{1}{L}\int_0^L dx\sqrt{1+\psi'(x)^2}e^{\kappa(\psi(x)-\psi_{\rm max})}\,,\label{g}
 \ee
 has been introduced for convenience.

Minimization of $H$ leads to the equilibrium condition
 \bb
  \delta\ell=\psi_{\rm max}+\kappa^{-1}\ln g\,, \label{ell}
 \ee
where $\delta\ell=\ell-\ell_\pi$ is the deviation in the interface height from that on a planar wall with $\ell_\pi=-\kappa^{-1}\ln(\delta
\mu/a\kappa)$.

Now, by setting $\psi(x)=A\cos(kx)$, the function $g$ takes the form
 \bb
 g(A,k)=\frac{1}{2\pi}\int_{-\pi}^{\pi} d\phi\sqrt{1+A^2k^2\sin^2\phi}\,e^{-\kappa A(1-\cos(\phi))}\,, \label{gsin0}
 \ee
 which can be further simplified using the saddle-point approximation by expanding the integrand around $\phi=0$ to get
 \bb
g(A,k)\approx\frac{1}{2\pi}\int_{-\infty}^{\infty} d\phi\sqrt{1+A^2k^2\phi^2}\,e^{\frac{-\kappa A}{2}\phi^2}\,. \label{gsin}
 \ee

From this, it follows that  $g(A,k)$ can be expressed in the scaling form
 \bb
g=A^{-\frac{1}{2}}\mathcal{G}\left(Ak^2\right)\,, \label{gweak}
 \ee
where the scaling function satisfies $\mathcal{G}(0)\approx(2\pi\kappa)^{-1}$. In the limit of strongly corrugated walls, where  $A^2k^2\gg1$, the
integral simplifies to
\begin{eqnarray}
g&\approx&\frac{Ak}{2\pi}\int_{-\infty}^{\infty} d\phi |\phi|\,e^{\frac{-\kappa A}{2}\phi^2}\,,
\end{eqnarray}
implying that $g\propto k$ with only a very weak $A$-dependence (due to neglected higher orders).

Clearly, the scaling form of $g$  given by Eq.(\ref{gweak}) breaks down in the $A\to0$ limit, where the saddle-point approximation leading to
Eq.(\ref{gsin}) becomes inappropriate. Instead, by expanding the integrand in Eq.~(\ref{gsin0}), we obtain the following approximation for small $A$:
 \bb
 g\approx1-\kappa A+\frac{A^2}{2}\left(k^2+3\kappa^2\right)\;\;\;(A\to0)\,. \label{gA0}
 \ee

Hence, the behavior of $\delta\ell$ can be summarized as follows::\\

i) For fixed $k$, the dependence of $\delta\ell(A)$ on the amplitude $A$ is initially quadratic, with $\delta\ell$ being positive and scaling as
$\propto\kappa^2+k^2/2$, as follows from substituting (\ref{gA0}) into (\ref{ell}). Beyond this region, when $A\lesssim \xi_b$, the dependence
becomes linear with a logarithmic correction:
 \bb
 \delta\ell(A)=A-c_A\ln(\kappa A)\,, \label{ell_A_weak}
 \ee
 where $c_A>0$.\\

ii) For fixed $A$, we expect a quadratic dependence of $\delta\ell(k)$ as
 \bb
\delta\ell(k)=c_0+c_kk^2\,,  \label{ell_k_weak}
 \ee
 with $c_0, c_k>0$. However, for strongly corrugated walls, we anticipate a crossover of $\delta\ell(k)$ to
  \bb
\delta\ell(k)=\tilde{c}_k\ln(k)\,,  \label{ell_k_strong}
 \ee
with $\tilde{c}_k>0$, implying the presence of an inflection point.

\subsubsection{Interface corrugation}

We now relax the assumption of a flat interface and investigate the decay of the interface corrugation as $\delta\mu\to0$.  Eq.~(\ref{omega11}) then
takes the form
 \begin{eqnarray}
 \Omega_1^1&=&\frac{\kappa}{2\pi}\int_0^L dx_1\sqrt{1+\psi'^2(x_1)}\int_0^L dx_2\sqrt{1+\ell'^2(x_2)}\cdot\nonumber\\
 &&\cdot\int_{-\infty}^\infty dy\frac{e^{-\kappa r}}{r}\,, \label{omega11_nonflat0}
 \end{eqnarray}
with $r=\sqrt{r_\perp^2+y^2}$ and $r_\perp=\sqrt{(x_2-x_1)^2+(\ell(x_2)-\psi(x_1))^2}$. Owing to the fast decay of the Yukawa term, we may
approximate $\sqrt{r_\perp^2+y^2}\approx r_\perp\left(1+\frac{y^2}{2r_\perp^2}\right)$, which allows us to integrate over $y$, yielding
 \begin{eqnarray}
\Omega_1^1&\approx&\sqrt{\frac{\kappa}{2\pi}}\int_0^L dx_1\sqrt{1+\psi'^2(x_1)}\int_0^L dx_2\sqrt{1+\ell'^2(x_2)}\cdot\nonumber\\
&&\cdot\frac{e^{-\kappa r_\perp}}{\sqrt{r_{\perp}}}\,. \label{omega11_nonflat}
 \end{eqnarray}

Either form of $\Omega_1^1$ can be used to determine  the equilibrium interfacial profile by solving  Eq.~(\ref{EL}) numerically. However, an
alternative, simpler approach involves approximating the interface profile using the parameterization
 \bb
  \ell(x)=\ell+\epsilon\cos(kx)\,. \label{ell_eps}
 \ee
Substituting this into Eq.~(\ref{ham}), the equilibrium profile is obtained by minimizing $H(\ell,\epsilon)$.

Furthermore, the ansatz in Eq.~(\ref{ell_eps}) allows for predicting the asymptotic behaviour of the interface corrugation. In the limit of $\delta
\mu\to0$, where we expect that $\ell\gg A\gg\epsilon$, it is straightforward to show that the minimization of $H$ leads to
 \bb
 \epsilon\sim e^{-\kappa\ell}\sim\delta \mu\,. \label{eps_dry}
 \ee

\subsection{Sharp-kink approximation} \label{ssec_ska}

If the wall-fluid potential is long-ranged, the exponentially decaying terms arising from the bulk entropy, we addressed in the previous section,
become negligible and the adsorption phenomena are dominated by the system energy. In this case, the effective interfacial Hamiltonian can be based
on the sharp-kink approximation \cite{dietrich, schick} for the fluid density near the wall of the local height $\psi(x)$:
 \bb
 \rho(x,z)=
\begin{cases}
   0\,, & z<\psi(x)\,,\\
    \rho_l\,,& \psi(x)<z<\ell(x)\,,\\
  \rho_g & z>\ell(x)\,, \label{ska}
\end{cases}
 \ee
which we adopt for an analysis of complete wetting; hence $\ell(x)$ now represents the local height of the liquid film adsorbed at the wall-gas
interface. Specifically, for our substrate model with $\psi(x)=A\cos(kx)$, we assume that
 \bb
\ell(x)=\ell+\epsilon\cos(kx)\,,
 \ee
 with $\epsilon\ll\ell$. Within this approximation, the excess grand potential over the bulk gas per unit length and period can be expressed as a
 function of the mean fluid interface $\ell$ and the corrugation parameter $\epsilon$:
  \bb
\Omega^{\rm ex}(\ell,\epsilon)=\delta\mu\Delta\rho\ell L+\gamma\int_0^L dx\sqrt{1+\epsilon^2k^2\sin^2(kx)}+W(\ell,\epsilon)\,, \label{om_lr}
 \ee
where the first term of the grand potential accounts for presence of the metastable liquid, the second term represents the liquid-gas interfacial
energy and the third term is the binding potential describing the interaction energy between wall/liquid and liquid/gas surfaces.

The binding potential (per unit length and period) is obtained by integrating  $V_w^{a}$ -- the attractive part  of the  wall potential $V_w(x,z)$ --
over the entire volume occupied by gas
 \bb
W(\ell,\epsilon)=-\Delta\rho\rho_w\int_0^L dx\int_{\ell(x)}^\infty dz V_w^{a}(x,z)\,. \label{W}
 \ee
Given the relation $\epsilon\ll\ell$, it is reasonable to approximate $V_w(x,z)\approx V_\pi(z-\psi(x))$, allowing us to evaluate the double integral
in Eq.~(\ref{W}) to get
 \bb
W(\ell,\epsilon)\approx\frac{A_HL\ell}{[\ell^2-(A-\epsilon)^2]^{\frac{3}{2}}}\,,
 \ee
 with the Hamaker constant
 \bb
  A_H=\frac{\pi\epsilon_w\sigma^6\rho_w\Delta\rho}{3}\,.
  \ee

Within the SKA, the equilibrium liquid-gas interfacial profile is obtained by minimizing the excess grand potential (\ref{om_lr}) within the subspace
of profiles given by (\ref{ska}). The minimization of $\Omega^{\rm ex}$ with respect to $\ell$ leads to the condition
 \bb
\delta\mu\Delta\rho=\frac{2A_H}{\ell^3}+\frac{6A_H(A-\epsilon)^2}{\ell^5}+\cdots \label{deltap}
 \ee
where the higher-order terms  (denoted by the ellipses) are neglected.  Next, we express the interface mean height $\ell$ as
 \bb
 \ell=\ell_\pi+\delta\ell \label{ell0}
 \ee
 by separating $\ell$ into the dominating contribution
 \bb
 \ell_\pi=\left(\frac{2A_H}{\delta \mu}\right)^{-\frac{1}{3}}\,,
 \ee
corresponding to the asymptotic behaviour of the interface height for a planar wall, and the perturbation $\delta\ell$.   Substituting from
(\ref{ell0}) into (\ref{deltap}), we obtain
  \bb
 \delta\ell=\frac{(A-\epsilon)^2}{\ell_\pi}\sim\delta \mu^{\frac{1}{3}}\,. \label{delta_ell}
  \ee

\begin{figure*}[tbp]
  \subfloatflex{ $L=20\,\sigma$, $A=2\,\sigma$}{%
    \includegraphics[scale=0.67]{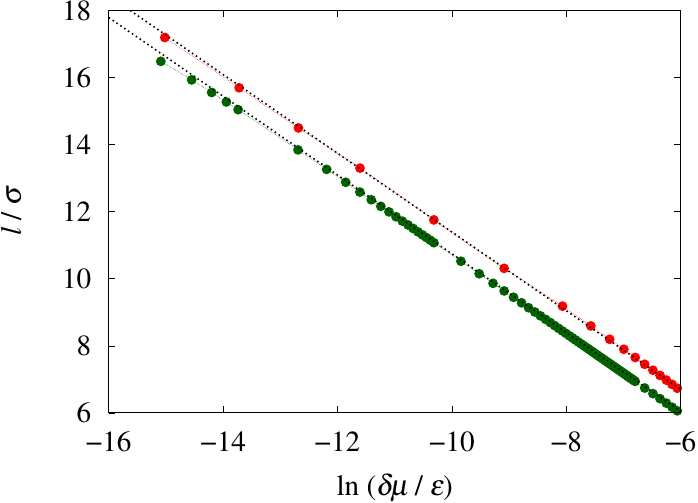}%
  } \hspace*{0.01cm}
  \subfloatflex{ $L=100\,\sigma$, $A=5\,\sigma$}{%
    \includegraphics[scale=0.67]{{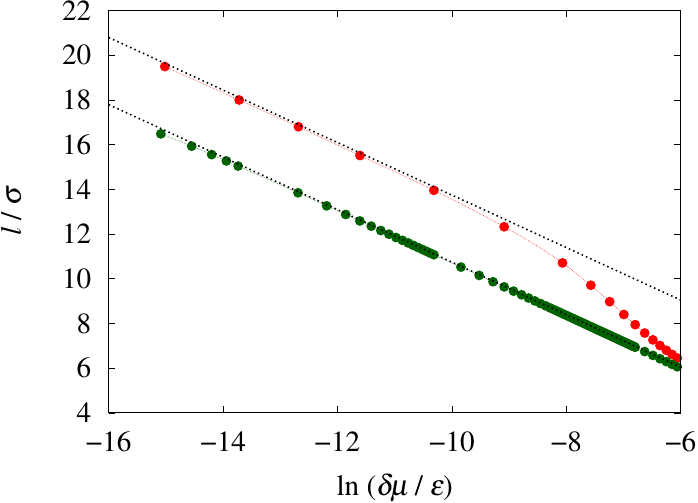}}%
  }
\caption{Semi-log plot comparing the growth of the interface height at sinusoidal walls, $\ell$ (red), to that at a planar wall, $\ell_\pi$ (green),
near saturation as obtained from DFT. The fitted lines have a slope of $-\xi_b$, confirming the expected logarithmic divergence of both $\ell$ and
$\ell_\pi$ in the limit $\delta\mu\to0$. The respective asymptotes are vertically shifted by  $\delta\ell=0.7\,\sigma$ for the wall with
$L=20\,\sigma$, $A=2\,\sigma$, and $\delta\ell=3\,\sigma$ for the wall with $L=100\,\sigma$, $A=5\,\sigma$, showing excellent agreement with the
prediction of Eq.~(\ref{ell}).} \label{ell_sr}
\end{figure*}

Furthermore, the minimization of $\Omega^{\rm ex}$ with respect to $\epsilon$ yields
  \bb
\epsilon=\frac{3A_H}{\gamma}\frac{\ell(A-\epsilon)}{k^2[\ell^2-(A-\epsilon)^2]^{\frac{5}{2}}}\sim\ell^{-4}\sim(\delta \mu)^{\frac{4}{3}}\,,
\label{eps_ska}
 \ee
indicating a somewhat faster decay than in the case of SR systems.

\section{Numerical results}

In this section, we test the scaling relations derived in the previous section against numerical DFT calculations. The comparison starts with an
examination of complete drying in SR systems, followed by a discussion of complete wetting in LR systems. Throughout our calculations, the
temperature has been fixed at $T=0.92\,T_c$,  ensuring the system remains above the wetting temperature of the attractive substrate model while
avoiding bulk criticality. All results are presented in reduced units, defined in terms of the microscopic parameters $\varepsilon$ and $\sigma$.

\subsection{Complete drying -- SR systems}


\begin{figure*}[tbp]
  \subfloatflex{}{%
    \includegraphics[scale=0.67]{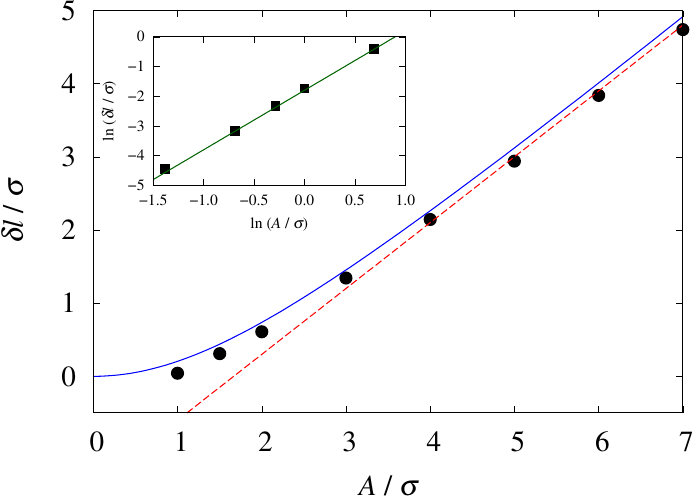}%
  } \hspace*{0.01cm}
  \subfloatflex{}{%
    \includegraphics[scale=0.67]{{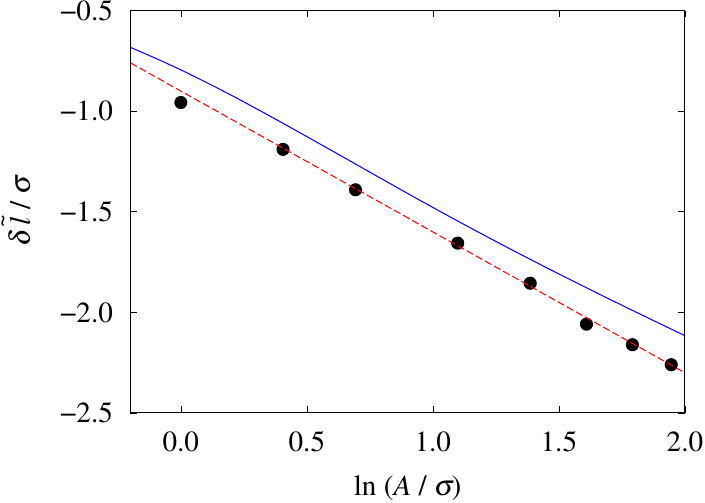}}%
  }
\caption{(a) The deviation of the interface height, $\delta\ell=\ell-\ell_\pi$, from its planar-wall value as a function of the wall amplitude $A$
for hard sinusoidal walls with the fixed period of $L=100\,\sigma$. The solid line represents the solution of the nonlocal Hamiltonian obtained by
solving numerically Eq.(\ref{Hflat}), while the symbols denote DFT results. The dashed line with slope 1 confirms the linear regime of
$\delta\ell(A)$. The inset shows a log-log plot, where the linear fit to the nonlocal Hamiltonian numerics with slope 2 supports the expected
quadratic regime of $\delta\ell(A)$ for small $A$. (b) Verification of the negative logarithmic correction to $\delta\ell$ [cf.
Eq.(\ref{ell_A_weak})] through the dependence of $\delta\tilde{\ell}=\delta\ell-A$ on $A$. The solid line corresponds to the numerical solution of
Eq.~(\ref{Hflat}), while symbols represent DFT results. The dashed line shows a linear fit to the DFT data.} \label{fixed-k-small-roughness-ell}
\end{figure*}

\begin{figure}[htbp]
  \includegraphics[width=\linewidth]{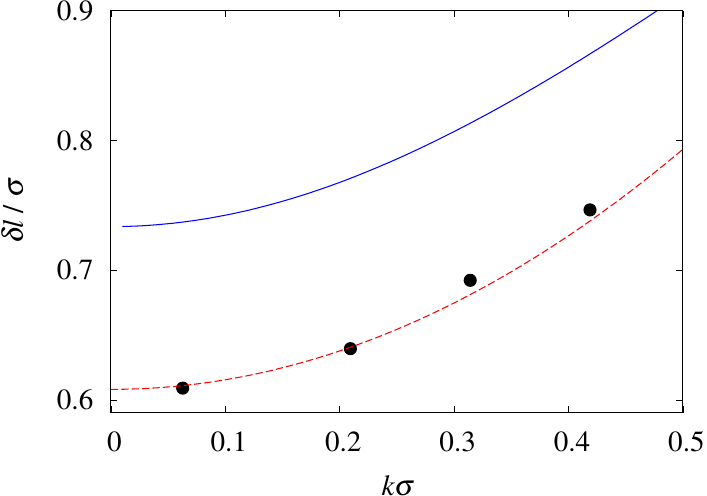}
  \caption{The deviation of the interface height, $\delta\ell=\ell-\ell_\pi$, from its planar-wall value as a function
of the wall wave number $k$ for hard sinusoidal walls with the fixed amplitude of $A=2\,\sigma$. The solid line represents the solution of the
interface Hamiltonian theory obtained by numerically solving Eq.(\ref{Hflat}), while symbols denote DFT results. The dashed line is a quadratic fit
to the DFT data confirming the expected dependence of $\delta\ell$ on $k$ [cf. Eq.(\ref{ell_k_weak})] for weakly corrugated walls, with the additive
constant $0.61\,\sigma$, which is slightly below the theoretical value $c_0\approx0.74\,\sigma$.}
  \label{fixed-A-small-roughness}
\end{figure}

\begin{figure}[hbtp]
  \subfloatflex{}{%
    \includegraphics[scale=0.67]{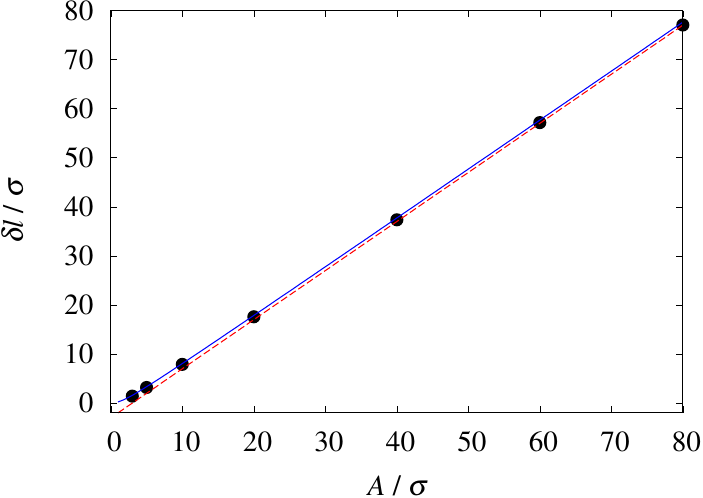}%
  } \\[1em]
  \subfloatflex{}{%
    \includegraphics[scale=0.67]{{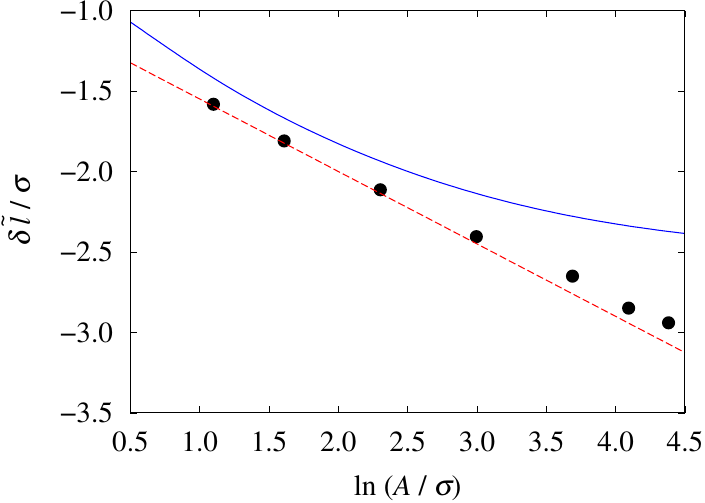}}%
  }
\caption{Same as in Fig.~\ref{fixed-k-small-roughness-ell},  but for strongly corrugated walls with period $L=20\,\sigma$ and large values of $A$. }
\label{fixed-k-large-roughness-ell}
\end{figure}

\begin{figure}[pht]
  \includegraphics[width=\linewidth]{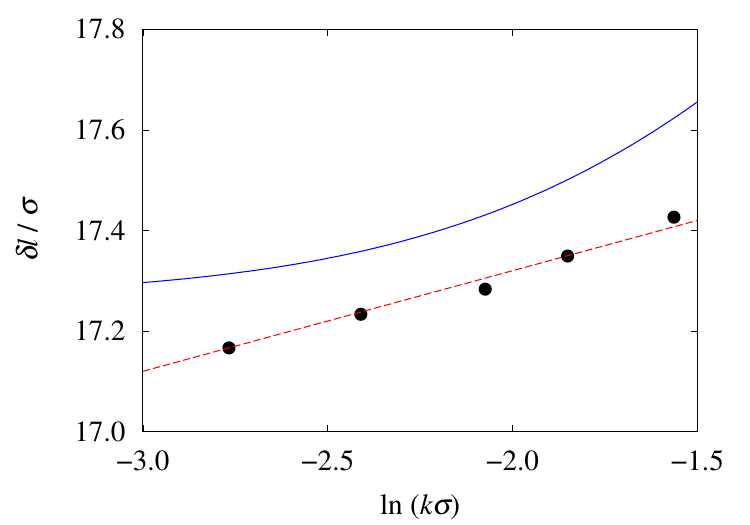}
  \caption{Semi-log plot confirming the logarithmic dependence of the interface height deviation, $\delta\ell=\ell-\ell_\pi$,
   on the wall wave number $k$, as expected for strongly corrugated hard walls [cf. Eq.~(\ref{ell_k_strong})].
  The solid line represents the solution of the interface Hamiltonian theory obtained by numerically solving Eq.(\ref{Hflat})
  and the dashed line shows a linear fit to the DFT data which are denoted by symbols. The wall amplitude  is fixed at $A=20\,\sigma$.}
  \label{fixed-A-large-roughness}
\end{figure}

\begin{figure}[pht]
  \includegraphics[width=\linewidth]{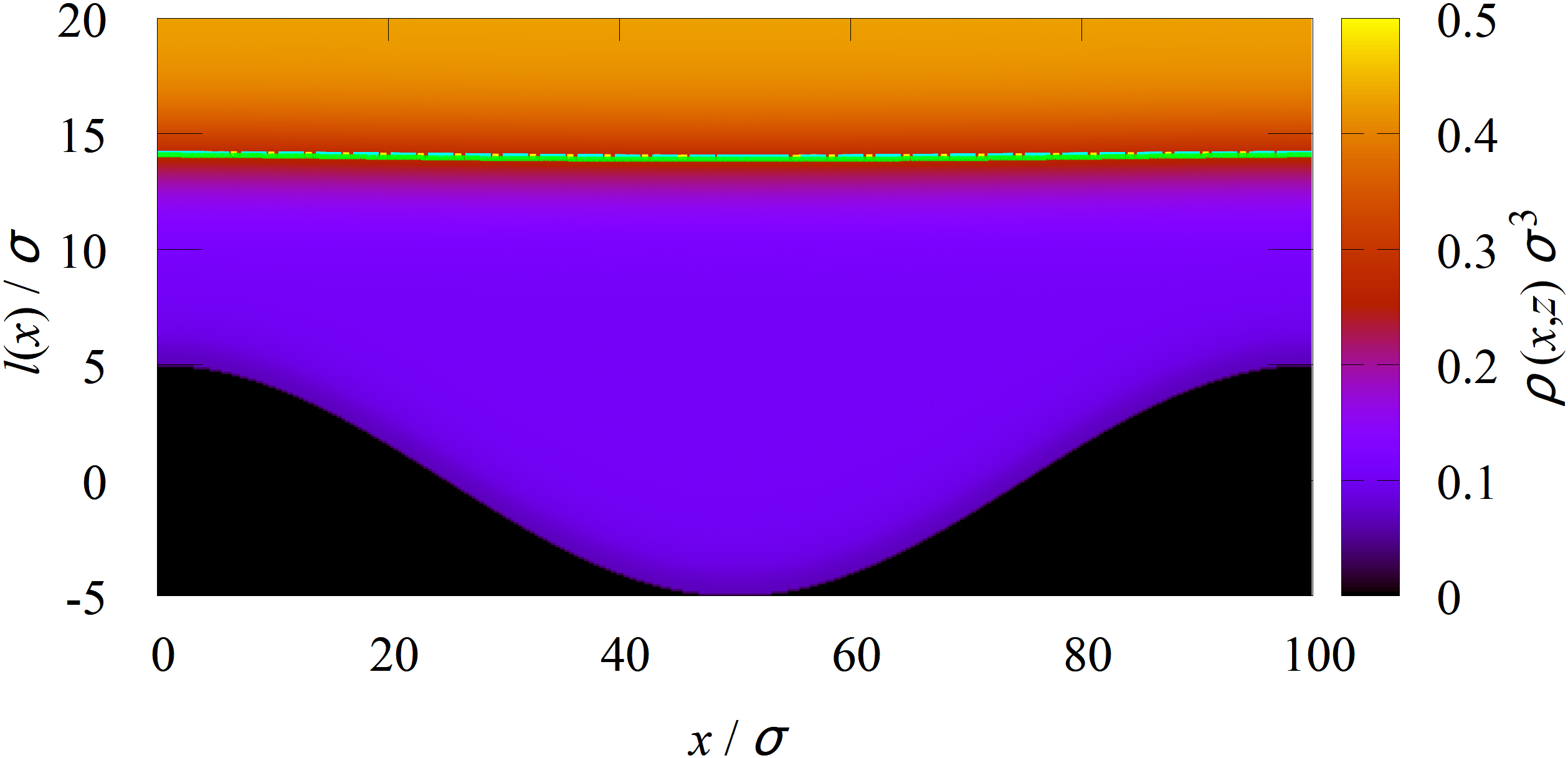}
  \caption{The equilibrium density profile of a fluid which in contact with a hard sinusoidal wall with $A=5\,\sigma$ and $L=100\,\sigma$ near saturation
   ($\delta\mu=10^{-5}\,\varepsilon$).  The solid line represents the liquid-gas interface obtained from DFT using the criterion in Eq.(\ref{dft_cont}),
    the dashed line corresponds to the interface profile from the nonlocal Hamiltonian with the parameterized form of $\ell(x)$ [Eq.(\ref{ell_eps})],
     and the dotted line depicts the interface obtained from the full minimization of the nonlocal Hamiltonian. In the current resolution, the
     three lines are almost indistinguishable, cf. Fig.~\ref{ifaces_L100_dft}.  }
  \label{dens_prof100}
\end{figure}

\begin{figure}[pht]
  \includegraphics[width=\linewidth]{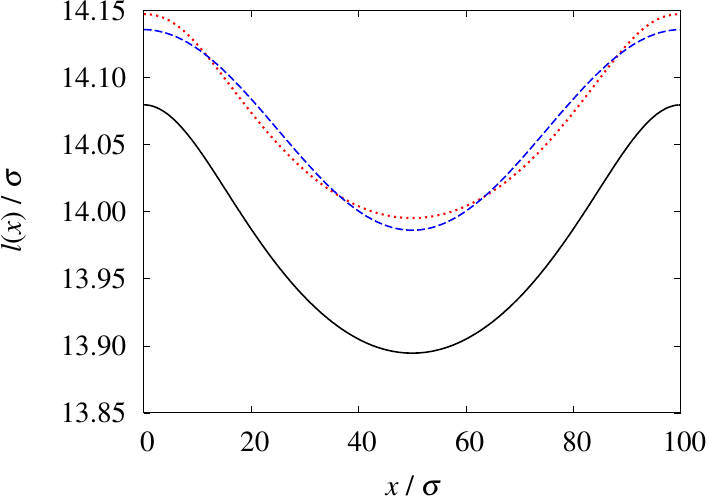}
  \caption{Detailed comparison of interface profiles obtained from DFT using the criterion in Eq.~(\ref{dft_cont}) (solid line),
  the approximation given by Eq.(\ref{ell_eps}) (dashed line), and the full functional minimization of the nonlocal Hamiltonian (dotted line).
  The results correspond to a hard sinusoidal wall with $A=5\,\sigma$ and $L=100\,\sigma$, and supersaturation $\delta\mu=10^{-5}\,\varepsilon$.}
  \label{ifaces_L100_dft}
\end{figure}

\begin{figure*}[p]
  \subfloatflex{$L=20\,\sigma$, $A=2\,\sigma$}{%
    \includegraphics[scale=0.67]{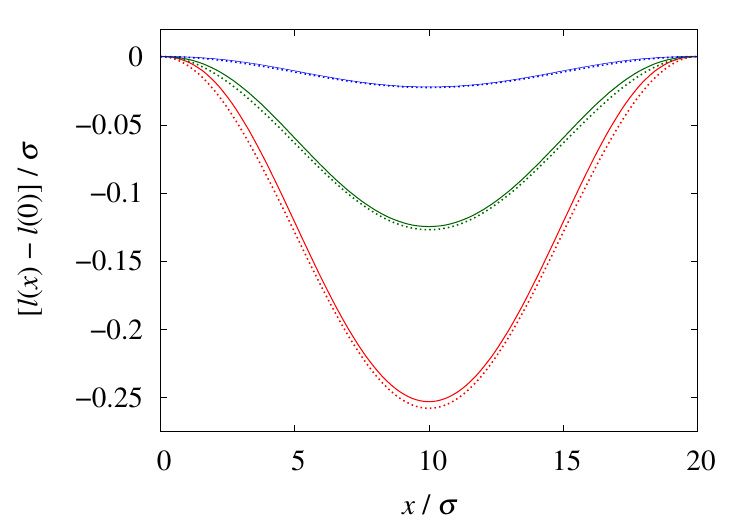}%
  }\hspace{0.01cm}%
  \subfloatflex{ $L=100\,\sigma$, $A=5\,\sigma$}{%
    \includegraphics[scale=0.67]{{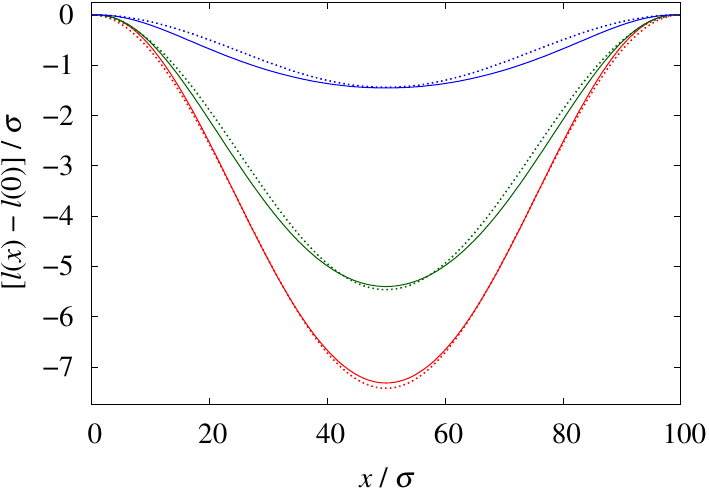}}%
  }
\caption{Comparison of equilibrium interface profiles, $\ell(x)$, as predicted by the nonlocal interface Hamiltonian theory. Solid lines show the
results from a full numerical minimization of the functional [Eq.(\ref{ham})], while dashed lines correspond to the approximate parameterized
solution given by Eq.(\ref{ell_eps}). Profiles are shown for three values of $\delta\mu/\varepsilon$ (from bottom to top): $10^{-4}$, $5 \times
10^{-4}$, and $10^{-3}$.} \label{nht_comp}
\end{figure*}

\begin{figure*}[p]
  \subfloatflex{ $L=20\,\sigma$, $A=2\,\sigma$}{%
    \includegraphics[scale=0.67]{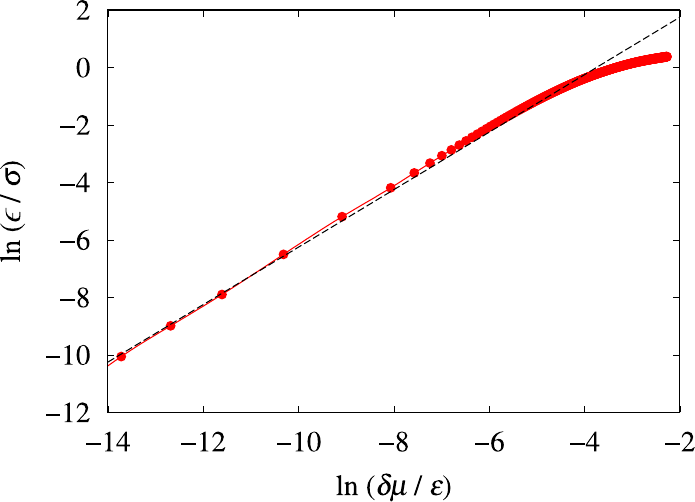}%
  }\hspace{0.01cm}%
  \subfloatflex{ $L=100\,\sigma$, $A=5\,\sigma$}{%
    \includegraphics[scale=0.67]{{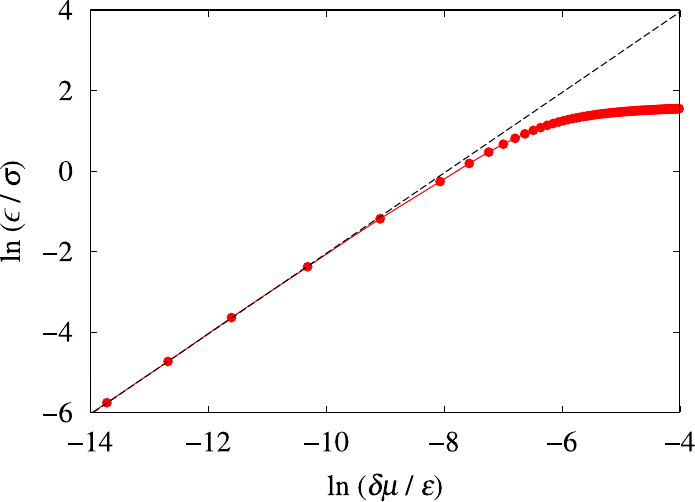}}%
  }
\caption{Log-log plot showing the expected linear dependence of the interface corrugation amplitude, $\epsilon$, on $\delta\mu$ for two illustrative
hard wall geometries. The solid line represents the solution of the interface Hamiltonian theory obtained by numerically solving Eq.(\ref{Hflat}) and
the dashed line shows a linear fit to the DFT data which are denoted by symbols. } \label{amplitude}
\end{figure*}

We begin by comparing the growth of the mean interface height, $\ell$, at sinusoidal walls in the limit  $\delta\mu\to0$  with that at a planar wall,
$\ell_\pi$. In Fig.~\ref{ell_sr} we show our DFT results for interface growth at two representative sinusoidal walls and the planar wall, plotted on
a semi-logarithmic scale. The results confirm the expected logarithmic divergence for all the walls with the same amplitude, which corresponds to the
(negative) bulk correlation length; the only effect of the wall's geometry is a vertical shift in the asymptotes. According to the nonlocal
Hamiltonian theory (NHT) prediction, as given by Eq.~(\ref{ell}),  the shift is $\delta\ell\approx0.8\,\sigma$ for the wall with $A=2\,\sigma$ and
$L=20\,\sigma$, and $\delta\ell\approx3.1\,\sigma$ for the wall with $A=5\,\sigma$ and $L=100\,\sigma$. These values are in excellent agreement with
the respective  DFT results, $\delta\ell\approx0.7\,\sigma$ and $\delta\ell\approx3.0\,\sigma$.

We now focus on testing the scaling properties of $\ell(A,k)$.
The presentation of our results is divided into two parts: first, we examine weakly corrugated walls,
followed by a discussion of strongly corrugated walls.\\

{\bf Weakly corrugated walls:} First, we investigate the dependence of $\ell$, or more precisely, the deviation $\delta\ell=\ell-\ell_\pi$, on the
wall amplitude $A$ for a fixed period of $L=100\,\sigma$. In Fig.~\ref{fixed-k-small-roughness-ell}a, we compare DFT results with NHT  solved in the
flat-interface approximation, showing excellent agreement across the entire range. The plot also confirms the expected crossover of $\delta\ell(A)$
from a quadratic dependence at small amplitudes to a linear regime at larger amplitudes.

The logarithmic correction to $\delta\ell(A)$, as predicted by Eq.(\ref{ell_A_weak}), is confirmed in the semi-logarithmic plot in
Fig.~\ref{fixed-k-small-roughness-ell}b. Here, the quantity $\delta\tilde\ell(A)=\delta\ell(A)-A$ exhibits a clear linear trend with the expected
negative slope. The agreement between DFT and NHT remains very satisfactory, despite a slight vertical shift between the respective sets of results.

Next, in Fig.~\ref{fixed-A-small-roughness}, we compare DFT and NHT results for $\delta\ell(k)$, this time fixing the amplitude at $A=2\sigma$ to
ensure small wall corrugations across the entire range. The logarithmic plot clearly confirms the expected quadratic growth of $\delta\ell(k)$ and
shows good quantitative agreement between the two approaches.

{\bf Strongly corrugated walls:} We now turn to strongly corrugated walls. First, we investigate the dependence of $\delta\ell(A)$ by reducing the
wall period to $L = 20\,\sigma$. As shown in Fig.~\ref{fixed-k-large-roughness-ell}a, $\delta\ell(A)$ exhibits an almost linear behavior, consistent
with the asymptotic prediction from Eq.~(\ref{ell_A_weak}), and shows excellent quantitative agreement between DFT and NHT. Furthermore, the
semi-logarithmic plot in Fig.~\ref{fixed-k-large-roughness-ell}b confirms the negative logarithmic correction to this dependence. Some deviations
between DFT and NHT are observed at very high amplitudes, which appear to mark the limits of the quantitative predictions of NHT. Next, in
Fig.~\ref{fixed-A-large-roughness}, we examine the behaviour of $\delta\ell(k)$ with a fixed amplitude of $A =20\,\sigma$. The DFT data confirm the
expected logarithmic dependence and show reasonable quantitative agreement with the NHT results.

Further, we examine the shape of the liquid-gas interface and its behaviour in the $\delta\mu \to 0$ limit. Figure~\ref{dens_prof100} shows an
illustrative two-dimensional equilibrium density profile for a sinusoidal wall with $A=5\,\sigma$ and $L=100\,\sigma$. The interface contours,
obtained from both DFT [Eq.~(\ref{dft_cont})] and NHT, are superimposed. For NHT, we employ both the full minimization of the nonlocal functional
$H[\ell]$ and the simplified parameterization given by Eq.~(\ref{ell_eps}). Notably, within the given resolution, the interface profiles are
virtually indistinguishable. To highlight this agreement, Fig.~\ref{ifaces_L100_dft} directly compares the three sets of contours $\ell(x)$,
demonstrating excellent consistency between NHT and DFT.

Next, we assess the accuracy of the parameterization (\ref{ell_eps}) by comparing it with the full numerical minimization of the functional
$H[\ell]$. Figure~\ref{nht_comp} presents the corresponding interface profiles for three representative states and two distinct wall geometries. We
observe a perfect consistency between the two approaches, particularly for the wall  $A = 2\,\sigma$ and $L = 20\,\sigma$, where the profiles
essentially overlap. This justifies the reliability of parameterization (\ref{ell_eps}) and indicates that its accuracy is more sensitive to the wall
amplitude than to the wall period.

Finally, we test the predicted asymptotic relation (\ref{eps_dry}) for the decay of interface corrugation. Figure~\ref{amplitude} displays DFT
results for the corrugation magnitude $\epsilon$, computed via Eq.~(\ref{eps_dft}), for two wall geometries down to very small values of $\delta\mu$.
In both cases, the DFT results converge convincingly to the respective asymptotic lines, confirming the expected scaling $\epsilon \sim \delta\mu$.

\subsection{Complete wetting -- LR systems}

\begin{figure}[tbhp]
  \includegraphics[width=\columnwidth]{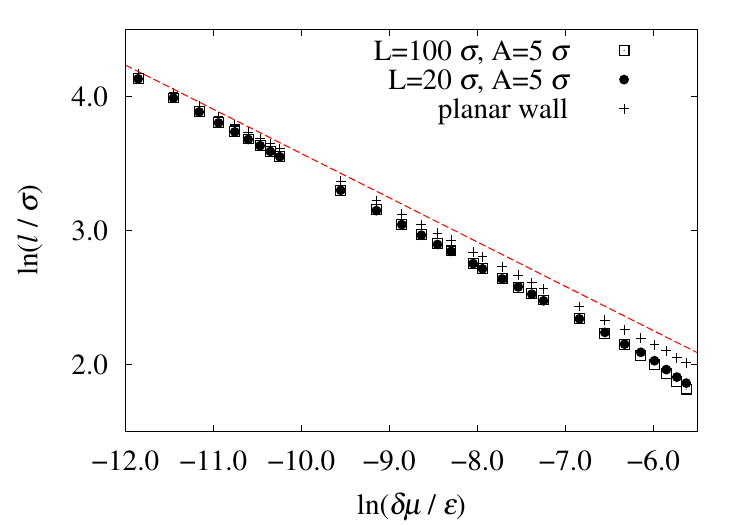}
  \caption{Comparison of interface growth for two illustrative sinusoidal models and a planar wall. In all cases, the data asymptotically
   approach the dashed line with a slope of $-1/3$, in agreement with the expected scaling behaviour $\ell\sim\delta\mu^{-1/3}$.}
  \label{ell_lr}
\end{figure}

\begin{figure}[htbp]
  \vskip-0.2em
  \subfloatflex{ $L=20\,\sigma$, $A=5\,\sigma$}{%
    \includegraphics[scale=0.67]{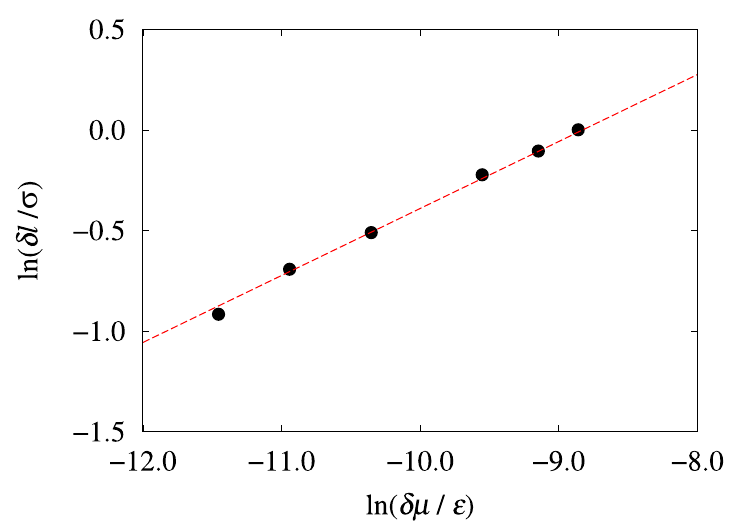}%
  } \\[0.5em]
  \subfloatflex{ $L=100\,\sigma$, $A=5\,\sigma$}{%
    \includegraphics[scale=0.67]{{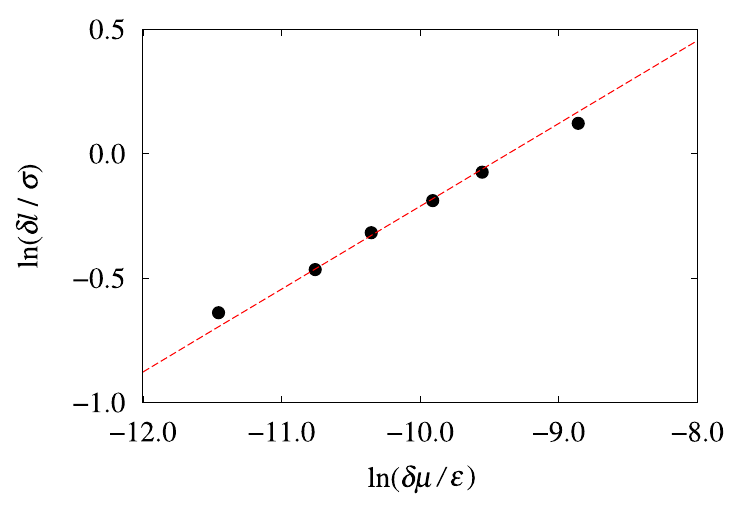}}%
  }
\caption{Log-log plot of the deviation of the interface height, $\delta\ell=\ell-\ell_\pi$, from its planar-wall value as a function of  $\delta\mu$
for attractive walls.  The symbols represent DFT results and the dashed line with slope $1/3$ confirms the expected power-law dependence
$\delta\ell\sim\delta\mu^{1/3}$ [cf. Eq.(\ref{delta_ell})]. } \label{delta_ell_lr}
\end{figure}

\begin{figure*}[pht]
  \subfloatflex{ $L=20\,\sigma$, $A=5\,\sigma$}{%
    \includegraphics[width=\columnwidth]{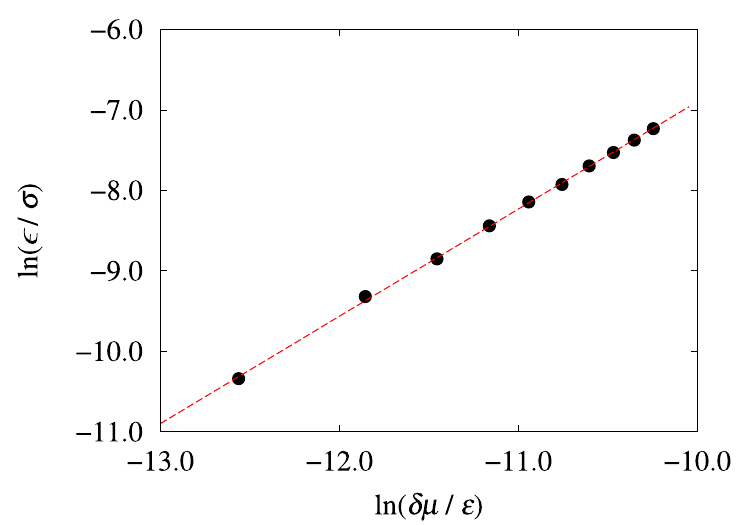}%
  } \hspace*{0.01cm}
  \subfloatflex{ $L=100\,\sigma$, $A=5\,\sigma$}{%
    \includegraphics[width=\columnwidth]{{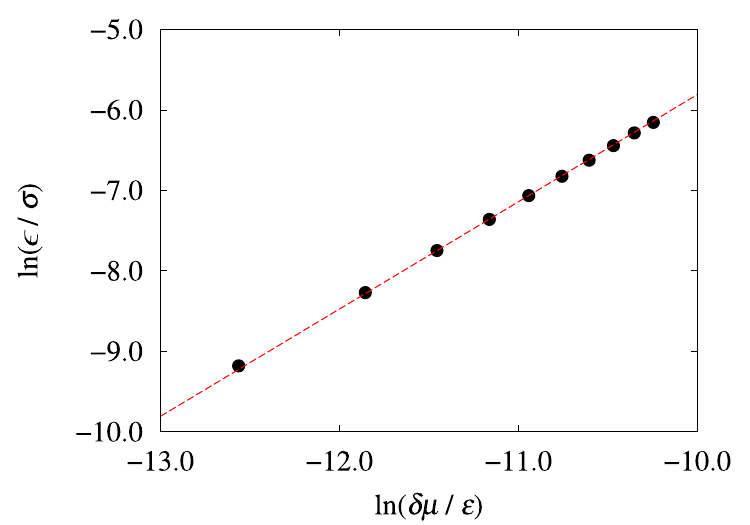}}%
  }
  \vskip-1ex
\caption{Log-log plot of  the interface corrugation amplitude, $\epsilon$, as a function of $\delta\mu$ for attractive walls. The symbols represent
DFT results and the dashed line with slope $4/3$ confirms the expected power-law dependence $\epsilon\sim\delta\mu^{4/3}$ [cf. Eq.(\ref{eps_ska})].}
\label{eps_lr}
\end{figure*}

\begin{figure*}[pht]
  \subfloatflex{ $L=20\,\sigma$}{%
    \includegraphics[width=\columnwidth]{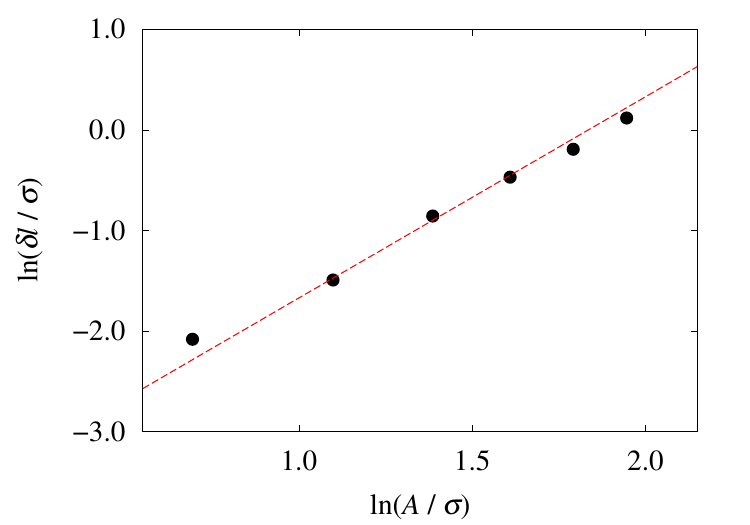}%
  } \hspace*{0.01cm}
  \subfloatflex{ $L=100\,\sigma$}{%
    \includegraphics[width=\columnwidth]{{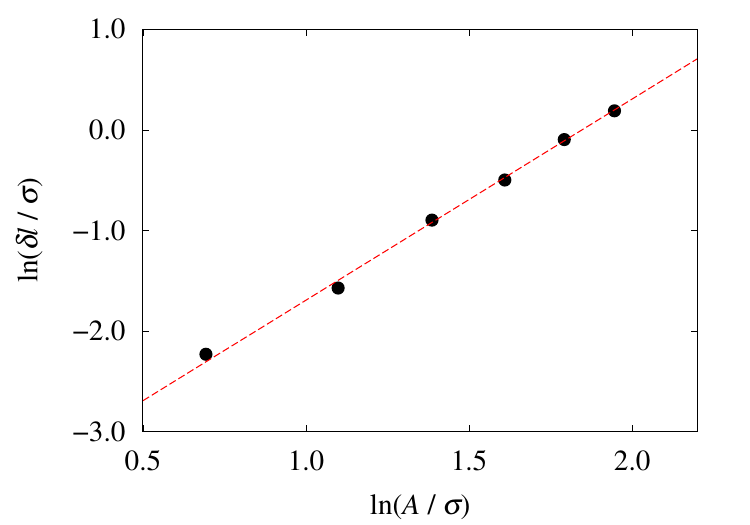}}%
  }
  \vskip-1ex
\caption{Log-log plot of the deviation of the interface height, $\delta\ell=\ell-\ell_\pi$, from its planar-wall value as a function of  the wall
amplitude $A$ for attractive walls.  The symbols represent DFT results and the dashed line with slope $2$ confirms the expected power-law dependence
$\delta\ell\sim A^2$ [cf. Eq.(\ref{delta_ell})].} \label{delta_ell_A_lr}
\end{figure*}

\begin{figure*}[pht]
  \subfloatflex{ $L=20\,\sigma$}{%
    \includegraphics[width=\columnwidth]{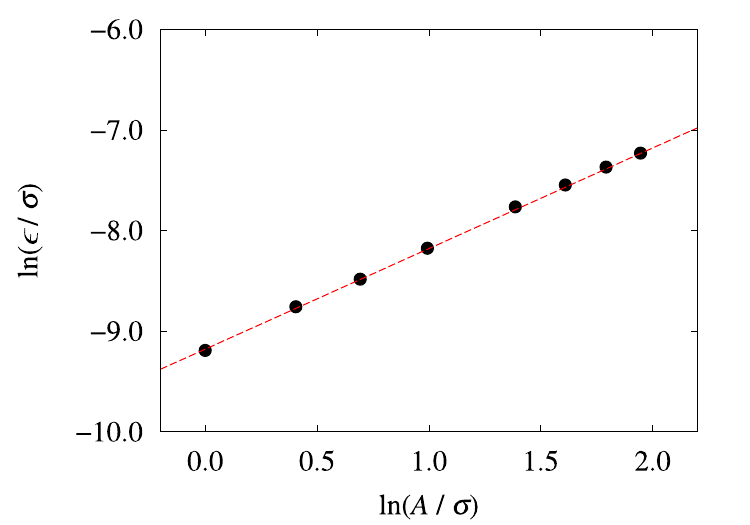}%
  } \hspace*{0.01cm}
  \subfloatflex{ $L=100\,\sigma$}{%
    \includegraphics[width=\columnwidth]{{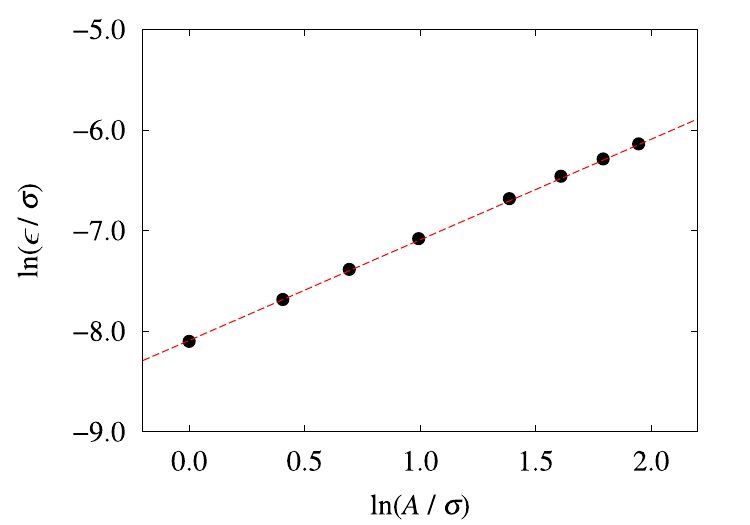}}%
  }
  \vskip-1ex
\caption{Log-log plot of  the interface corrugation amplitude, $\epsilon$, as a function of  the wall amplitude $A$ for attractive walls. The symbols
represent DFT results and the dashed line with slope $1$ confirms the expected linear dependence  [cf. Eq.(\ref{eps_ska})].} \label{eps_A_lr}
\end{figure*}

We now present our DFT results for complete wetting at sinusoidal walls exerting a long-ranged potential, aiming to test the scaling predictions
based on SKA,  as described in Sec.~\ref{ssec_ska}. Unlike in SR systems, no corrugation-dependent regimes are expected here. Nevertheless, we
present our results by examining two classes of systems with varying levels of roughness, namely: (i) $L=20\,\sigma$ and (ii) $L=100\,\sigma$. We
begin by presenting our DFT results in Fig.~\ref{ell_lr}, showing the dependence of $\ell(\delta\mu)$ for both substrate models with $A=5\,\sigma$.
In both cases, the results exhibit the expected power-law scaling with the critical exponent $\beta^{\rm co}=1/3$, i.e. the same as for a flat wall,
whose results are also included for comparison.  To highlight subtle corrections to the leading $(\delta\mu)^{-1/3}$ dependence of $\ell(\delta\mu)$,
we present in Fig.~\ref{delta_ell_lr}  the deviation $\delta\ell(\delta\mu)$ of the wetting layer thickness from that on a planar wall.   In full
agreement with the SKA analysis, the DFT results conclusively confirm the predicted scaling $\delta\ell \sim \delta\mu^{1/3}$ as $\delta\mu$ tends to
zero.

Next, we complement our examination of the interface growth by analyzing the decay of interface corrugations. In Fig.~\ref{eps_lr} we present DFT
results showing the $\delta\mu$-dependence of $\epsilon$, the interface amplitude.  The logarithmic plots reveal the asymptotic power-law behaviour
$\epsilon \sim \delta\mu^{4/3}$, in full agreement with the SKA prediction given by Eq.(\ref{eps_ska}).

Finally, we test our predictions for the dependence of interface behaviour on wall amplitude $A$. Fig.~\ref{delta_ell_A_lr} illustrates how
$\delta\ell$ changes with $A$ for different fixed values of $\delta\mu$. The corresponding log-log plots exhibit excellent agreement with the
expected linear dependence, with a slope of $2$ (cf. Eq.~(\ref{delta_ell})), particularly for weakly corrugated walls. Likewise, the DFT results
shown in Fig.~\ref{eps_A_lr} strongly support the predicted linear dependence of $\epsilon$ on $A$.

\section{Conclusion}

We have investigated the impact of wall corrugation on the critical phenomena of complete wetting and drying, characterized by the divergence of the
adsorbed layer thickness as saturation is approached. The substrate was modelled as a sinusoidally corrugated surface in one spatial direction with
amplitude $A$ and wave number $k$. Our objective was to contrast the properties of the liquid-gas interface -- separating the adsorbed layer from the
bulk fluid -- with those observed at a planar substrate.

Our study involved two distinct scenarios: (i) complete drying in systems with only short-ranged (SR) interactions, modelled by a purely repulsive
hard wall, and (ii) complete wetting involving long-ranged (LR) forces, where the microscopic model consisted of a wall of uniformly distributed
Lennard-Jones atoms. To analyze these systems, we employed different mesoscopic theoretical approaches suited to their respective interaction ranges.
For SR interactions, we applied the nonlocal Hamiltonian theory, adapted for interfacial phenomena in general geometries. For LR interactions, we
used the sharp-kink approximation, allowing us to determine the effective binding potential for weakly corrugated walls. In both cases, we derived
scaling relations describing the interface height and shape in the asymptotic limit $\delta\mu\to0$, which can be summarized as follows:

\begin{itemize}

\item {\bf For SR systems}, the deviation of the interface height, $\delta\ell$, from a planar wall depends solely on the geometric wall parameters.
For small amplitudes  $\delta\ell$ scales quadratically with $A$ but crosses over to a linear regime with a negative logarithmic correction as the
amplitude becomes comparable to the bulk correlation length. For a fixed $A$, $\delta\ell(k)$ initially grows quadratically with $k$ but becomes
logarithmic in the strong corrugation regime ($A^2k^2\gg1$). The amplitude of the interface corrugation vanishes linearly with $\delta\mu$ as the
system approaches saturation.

\item {\bf For LR systems} with dispersion forces,  $\delta\ell$ includes a substantial contribution from $\delta\mu$,
scaling as $\delta\ell\sim (\delta\mu)^{1/3}$, and grows quadratically with $A$. The amplitude of the interface corrugation scales as
$(\delta\mu)^{4/3}$ and increases linearly with  $A$.

\end{itemize}

These predictions were validated through fully microscopic density functional theory (DFT) calculations, which confirmed the scaling relations,
including those for strongly corrugated walls in SR systems. Moreover, the agreement between DFT and the (mesoscopic) nonlocal Hamiltonian theory was
quantitatively surprisingly accurate; to our knowledge, this is the first microscopic test of this theory.

We emphasize that the key differences between the two models arise not from whether the adsorbed phase is liquid or gas, but rather from the nature
of the underlying microscopic forces. Regarding the generality of our results for LR systems, our conclusions apply specifically to (non-retarded)
dispersion forces, where intermolecular interactions decay as $r^{-(p+4)}$ with $p=2$. A natural question arises: how do these results change with
$p$, given that for a planar wall, the asymptotic form of the binding potential follows $W_\pi \sim \ell_\pi^{-p}$, with critical exponents
$\beta^{\rm co} = 1/(p+1)$ and $\nu_\parallel^{\rm co}=(p+2)/(2p+2)$? It is straightforward to show, using essentially just dimensional arguments,
that the scaling behaviour of $\delta\ell$ remains unchanged, confirming its universality within LR systems. However, while the interface corrugation
amplitude retains the same dependence on wall parameters, its power-law decay with $\delta\mu$ is modified to
$\epsilon\sim(\delta\mu)^{-\frac{p+2}{p+1}}$, suggesting that $\epsilon\propto\left(\nu_\parallel^{\rm co}\right)^{-2}$.

In summary, our study reveals how wall corrugation influences the scaling properties of complete wetting and drying, with a particular focus on the
significance of the nature of microscopic interactions. We demonstrated that the nonlocal Hamiltonian theory and the sharp-kink approximation offer
complementary analytic frameworks for describing wetting phenomena on corrugated surfaces in SR and LR systems, respectively. These approaches enable
the formulation of scaling relations for both the growth and the shape of the unbinding interface -- predictions that are in excellent agreement with
fully microscopic DFT calculations.

Future research can build upon these findings by exploring more intricate and realistic surface geometries, such as randomly rough substrates, hierarchical or fractal structures, chemically patterned interfaces, and substrates exhibiting mixed-scale features. Investigating these systems would not only test the robustness and limitations of the current analytical approaches but also potentially reveal new regimes of geometric scaling and fluctuation-induced phenomena that are not captured by simpler sinusoidal models.

Moreover, the study of adsorption on complex geometries may be further extended to non-equilibrium systems, especially those involving self-propelled
particles. The wetting behavior of such active matter has received growing attention in recent literature \cite{neta21, turci21, turci24}, and
exploring its interaction with non-planar surface topographies represents an intriguing challenge
 for future research.

Finally, while our present study has focused on mean-field descriptions of interface profiles, it would be valuable to assess the role of
thermal interfacial fluctuations, which have been neglected in the current treatment. Although we do not expect such fluctuations to qualitatively
alter the leading-order scaling behavior in three dimensions where their influence is typically subdominant, they may introduce quantitative
corrections or affect crossover regimes, especially near criticality or in low-dimensional systems. Incorporating these fluctuations, perhaps via
renormalization group approaches or simulations, would offer a more complete theoretical understanding of the wetting behaviour on structured
surfaces.

\begin{acknowledgments}
\noindent We are grateful to Andrew Parry for insightful discussions. The work was financially supported by the Army Research Office and was
accomplished under Cooperative Agreement No. W911NF-24-2-0244.
\end{acknowledgments}

\end{document}